\documentclass[manuscript]{aastex}
\usepackage{amsmath,amssymb}
\usepackage{graphicx}

\slugcomment{Accepted for Publication in Icarus: 27 October, 2015}

\shorttitle{Melting of Hadean Continental Crust by LHB}
\shortauthors{Shibaike, Sasaki & Ida}

\begin{document}

\title{Excavation and Melting of the Hadean Continental Crust \\ by Late Heavy Bombardment}

\author{Yuhito Shibaike}
\affil{Department of Earth and Planetary Sciences, Tokyo Institute of Technology, 2-12-1 Ookayama, Meguro-ku, Tokyo 152-8551, Japan}
\email{shibaike.y@geo.titech.ac.jp}

\author{Takanori Sasaki}
\affil{Department of Astronomy, Kyoto University, Kitashirakawa-Oiwake-cho, Sakyo-ku, Kyoto 606-8502, Japan}

\and

\author{Shigeru Ida}
\affil{Earth-Life Science Institute, Tokyo Institute of Technology, 2-12-1 Ookayama, Meguro-ku, Tokyo 152-8550, Japan}

\begin{abstract}
No Hadean rocks have ever been found on Earth's surface except for zircons---evidence of continental crust, suggesting that Hadean continental crust existed but later disappeared. One hypothesis for the disappearance of the continental crust is excavation/melting by the Late Heavy Bombardment (LHB), a concentration of impacts in the last phase of the Hadean eon. In this paper, we calculate the effects of LHB on Hadean continental crust in order to investigate this hypothesis. Approximating the size-frequency distribution of the impacts by a power-law scaling with an exponent $\alpha$ as a parameter, we have derived semi-analytical expressions for the effects of LHB impacts. We calculated the total excavation/melting volume and area affected by the LHB from two constraints of LHB on the moon, the size of the largest basin during LHB, and the density of craters larger than 20 km. We also investigated the effects of the value of  $\alpha$. Our results show that LHB does not excavate/melt all of Hadean continental crust directly, but over 70\% of the Earth's surface area can be covered by subsequent melts in a broad range of $\alpha$. If there have been no overturns of the continental crust until today, LHB could be responsible for the absence of Hadean rocks because most of Hadean continental crust is not be exposed on the Earth's surface in this case. 
\end{abstract}

\keywords{Earth --- Impact processes --- Prebiotic environments --- Asteroids}

\section{Introduction} \label{Introduction}

Hadean rocks have not been found on Earth until today, and the age of the oldest rock is about 4.0 Ga. No continental crust may have existed on the Hadean Earth. However, some zircons that formed during the Hadean eon were found in Jack Hills sedimentary rocks \citep{wil01}. Since zircons are formed by igneous activity at the same time as granite, this discovery suggests that the Hadean continental crust existed but later disappeared. The disappearance of the continental crust could be accounted for by geological activity such as reworking or plate tectonics, and erosion of the crusts \citep[e.g.,][]{kaw09}. Another possibility is excavation/melting of the continental crust by the Late Heavy Bombardment (LHB), a concentration of impacts considered to have existed in the last phase of the Hadean eon. In this paper, we calculate the effects of LHB on Hadean continental crust in order to investigate this hypothesis.

\subsection{Late Heavy Bombardment} \label{lhb}

A classic scenario of LHB has been based on the fact that radiometric dates of lunar basins' impact melts were concentrated at 3.9 Ga \citep{ter74}. In this model, about 15 lunar basins are considered to be formed between 3.9 and 3.8 Ga \citep[e.g.,][]{ryd02}. \citet{coh05} argued from Ar-geochronology of lunar meteorites that the concentration at 3.9 Ga is not required and \citet{har03} claimed that the decrease of lunar crater density is just the tail of the planetesimal accretion. The most recent model proposed a sawtooth-like timeline of impact flux \citep{mor12}. If LHB was caused by a disturbance in the ancient main belt asteroids, the impact flux decreased exponentially after the onset of LHB \citep{bot12}. This exponential curve fits the lunar crater density curve very well if the disturbance occurred at 4.1 Ga \citep{mor12}. It is adjusted in order to explain the accreted mass of highly siderophile elements (HSE) of the moon after the moon formation giant impact. \citet{gom05} argued that the cause of LHB was the migration of Jupiter and other giant planets predicted in the Nice model. The migration of the giant planets moved some resonances and scattered the main belt asteroids and outer comets.

\subsection{Previous works} \label{previous}

The estimate for coverage of the Earth's surface by impact craters depends on the impactors' size-frequency distribution (SFD). The estimates with unconstrained SFD predicted diverse results \citep[e.g.,][]{fre77, fre80, gri80, ryd00}. Furthermore, stochastic one or two huge impacts could give considerable effects \citep{zah97}.

Recently, using a more constrained SFD corresponding to the current main belt asteroids with the total LHB mass of $2\times10^{23}$ g, \citet{abr13} showed that LHB may not have melted all of Hadean continental crust. They computed 3D temperature distributions of the crust using an analytical shock-heating model with effects of impact melt generation, uplift, and ejecta heating. The result is that 1.5--2.5\% of the upper 20 km of the crust was melted during LHB, and only 0.3--1.5\% was melted through LHB period. They also indicated that 5--10\% of the Earth's surface area was covered by over 1 km depth of impact melt sheets, and the entire surface was covered by impact ejecta close to 1 km deep.

On the other hand, \citet{mar14} suggested that Hadean impacts could explain the absence of early terrestrial rocks based on the sawtooth-like timeline, with the current main belt asteroids' SFD. The key point of \citet{mar14} is the ``stochastic'' nature of LHB. They showed that the melting volume deeply depended on whether very large impactors hit Earth or not. Another key point is the effect of ``impact-induced decompression'' and subsequent adiabatic melting of rising material in the mantle, which increases the total melting volume. They claimed that these flood melts from under the crust emerge when the impactors' diameters were greater than 100 km. The melts flowed on the Earth's surface like spherical caps. The melt sheets' diameters were about 20--30 times as large as that of the impactors, for an assumed melt thickness of 3 km. The result is that 70--100\% of the Earth's surface area was covered by the melt sheets since 4.15 Ga, and 400--600\% was covered during the period 4.5--4.15 Ga. This value of the total melting area on the Earth's surface is about ten times as large as that of \citet{abr13}.

\subsection{Aims of this work} \label{ourwork}

Because these previous works were carried out based on respective model of the impactor's SFD and the effects of impact, the results were inconsistent with each other. This work is going to investigate the effects of the slope of the impactor's SFD, which is the main new contribution of this work.

\begin{figure}[tbsp]
\begin{center}
\includegraphics[width=15cm]{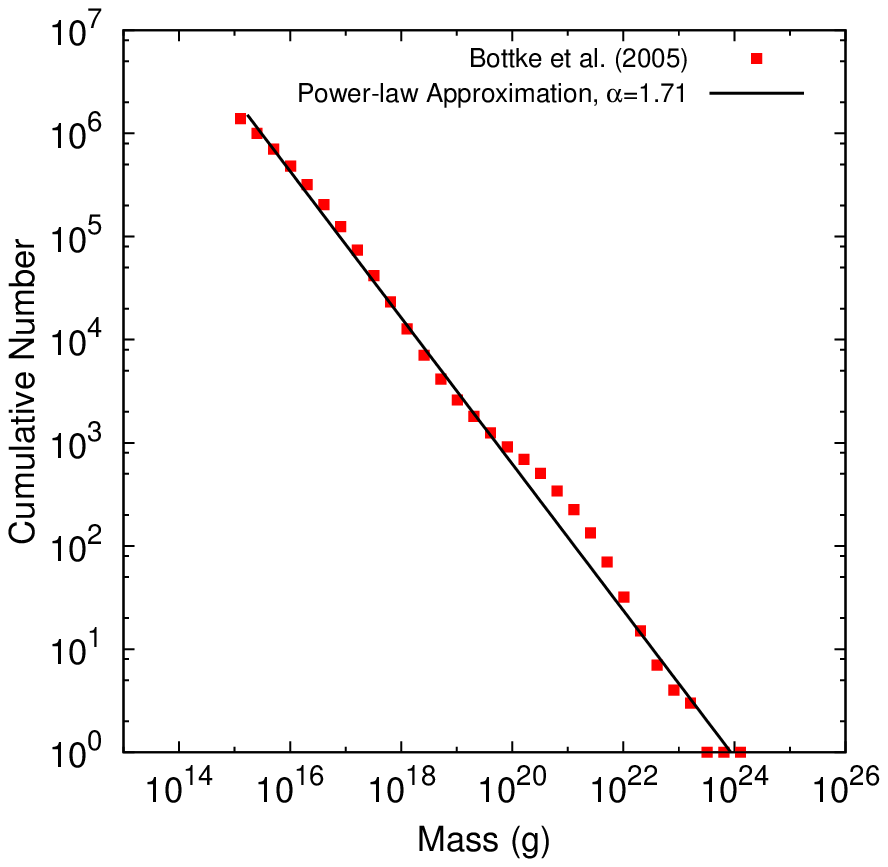}
\caption{Current main belt asteroids' SFD and its power-law approximation \newline Red dots are derived from observations \citep{bot05}. The bin width increases by a factor of 0.5. The black line is a power-law approximation of these plots and $\alpha =1.71$. We transformed the SFD into the mass-frequency distribution as the asteroids' densities are $\rho_{\rm i}=2.6$ g/cm$^{3}$; the value of density is consistent with the total mass of the main belt asteroids \citep{kra02}.} \label{fig1}
\end{center}
\end{figure}

 The SFD of LHB impactors is still controversial. Some studies claim that the lunar craters' SFD is consistent with the current main belt asteroids' SFD \citep[e.g.,][]{str05}. Figure \ref{fig1} shows that the current main belt asteroids' SFD can be approximated as $\alpha=1.71$, where (minus) $\alpha$ is the power index of the mass-frequency distribution \citep{bot05}. On the other hand, analytical theory of the evolution of SFD in a collision cascade predicted that $\alpha=1.83$ \citep{tan96}.

Corresponding to these uncertainties in the LHB impactors' SFD, we here consider a wide variety of SFDs. In order to reveal intrinsic physics more clearly, we approximate the SFD as a power-law function with a broad range of the power index $\alpha$, rather than adopting a single more detailed SFD.
The number of impactors heavier than $m$, $N_{\rm sfd}(>m)$, is defined as
\begin{equation}
\frac{\mathrm{d} N_{\rm sfd}}{\mathrm{d}m}=Am^{-\alpha},
\label{dN(m)}
\end{equation}
where $A$ is a proportional constant. Then,
\begin{eqnarray}
N_{\rm sfd}\!\!\!\!&=&\!\!\!\!\int_{m}^{\infty} Am'^{-\alpha}dm' \nonumber \\
&=&\!\!\!\!\frac{A}{\alpha-1}m^{1-\alpha},
\label{N(m)}
\end{eqnarray}
when $\alpha>1$ \citep{doh69}.

We estimate the effects of LHB in following way. In Section 2, we derive the semi-analytical expressions for the effects of LHB impacts. We estimate the total excavation/melting volume and area by integrating the effects of individual impacts, assuming a power-law SFD. In Section 3, we evaluate total effects with a fixed total mass of impactors and the two lunar constraints---the maximum size of basin formed during LHB and the small craters' density. We also investigate the dependence of the power index $\alpha$ of SFD. In Section 4, we discuss whether LHB can explain the absence of Hadean rocks. Then, we compare our results with those of the previous works for a nominal value of $\alpha =1.61$. We also discuss the validity of our assumptions and models. In Section 5, we summarize this paper. Appendix \ref{appendix} is devoted to the calculation of the value of $\alpha$ which fulfills both the lunar constraints.

\section{Basic Methods}

\subsection{Effects of a single impact} \label{singlei}

\begin{figure}[tbsp]
\begin{center}
\includegraphics[width=15cm]{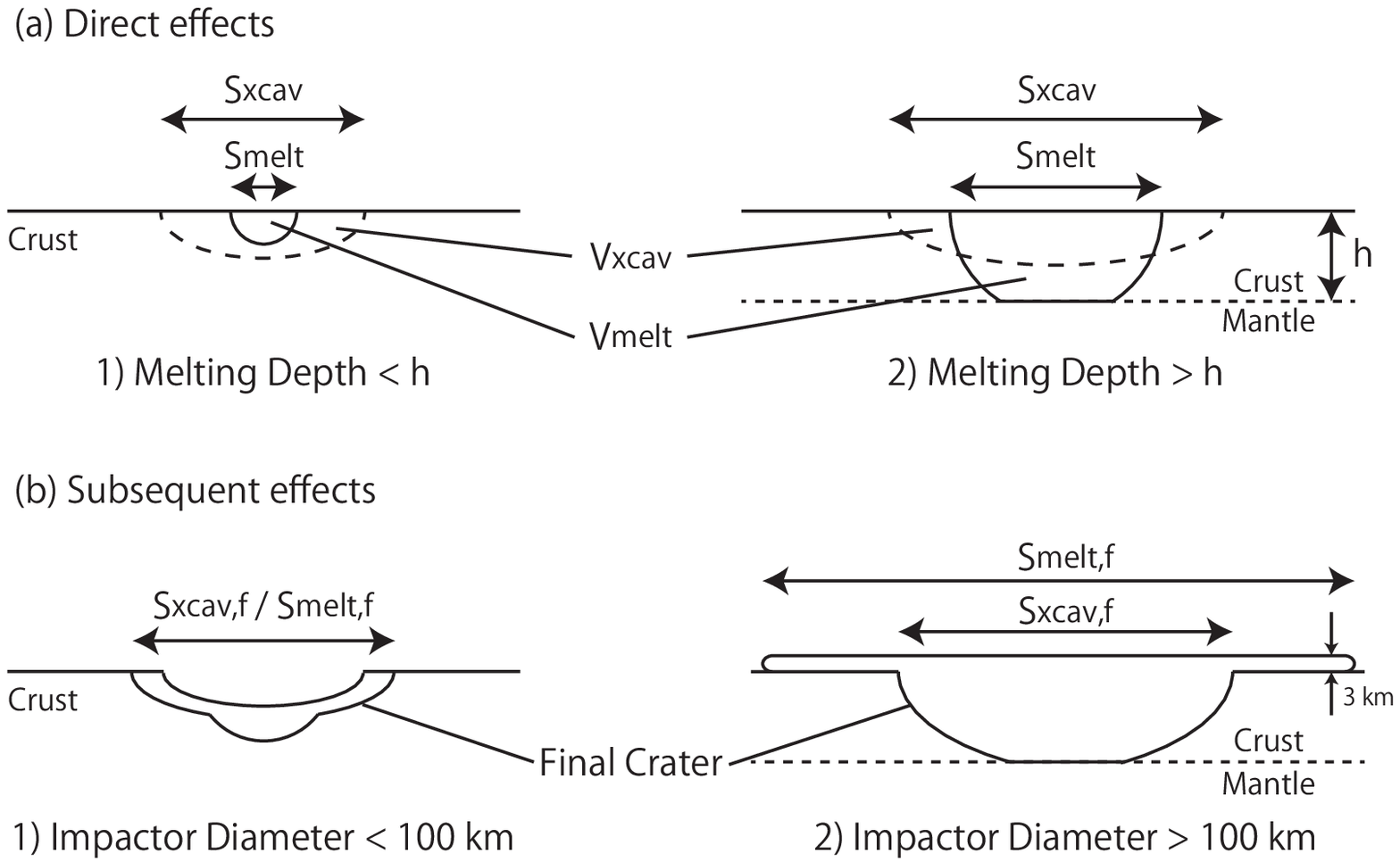}
\caption{Vertical cross sections of single impact effects \newline Excavation regions are framed by dashed curves, and melting regions are framed by solid curves. We defined four direct effects, ``direct excavation volume, $V_{\rm xcav}$,'' ``direct melting volume, $V_{\rm melt}$,'' ``direct excavation area, $S_{\rm xcav}$,'' and ``direct melting area, $S_{\rm melt}$.'' $S_{\rm xcav}$ and  $S_{\rm melt}$ are the horizontal cross sections of the direct excavation/melting regions. When the depth of the melting region is deeper than the thickness of the crust, we cut the volume beneath the crust. We also defined two subsequent effects, ``excavation area including gravitational collapse, $S_{\rm xcav,f}$" and ``melting area including the area covered by melts, $S_{\rm melt,f}$.'' $S_{\rm xcav,f}$ is the area of the final crater. The rim of the transient crater collapses with gravity and the final crater is formed. $S_{\rm melt,f}$ includes both the direct melting area and the area covered by melts. The direct melting area is surrounded by the area covered by melts. When the diameter of the impactor is smaller than 100 km, $S_{\rm melt,f}$ is equal to $S_{\rm xcav,f}$. When the diameter is larger than 100 km, $S_{\rm melt,f}$ is calculated by Eq. (\ref{Smeltf}).} \label{fig2}
\end{center} 
\end{figure}

We consider a single impact causes direct excavation/melting and subsequent excavation/melting. Formation of transient crater and melting by shock heating cause ``direct excavation/melting,'' while formation of final crater and uplifted or excavated molten rock spreading on and beyond the final crater cause `` subsequent excavation/melting,'' respectively. In this work, these effects are denoted $V_{\rm xcav}$, $V_{\rm melt}$, $S_{\rm xcav}$ and $S_{\rm melt}$ for direct effects, and $S_{\rm xcav,f}$ and $S_{\rm melt,f}$ for subsequent effects (Fig. \ref{fig2}). The volume of evaporated rocks would be sufficiently small so that we did not include it in this study.

We assume the direct excavation volume, $V_{\rm xcav}$, is the volume of the transient crater \citep{abr12},
\begin{eqnarray}
V_{\rm xcav}\!\!\!\!&=&\!\!\!\!0.146\left(\frac{gL}{v^2}\right)^{-0.66}\biggl(\frac{\rho_{\rm t}}{m}\biggr)^{-1}\mathrm{sin}^{1.3}\theta \nonumber \\
&=&\!\!\!\!0.127\rho_{\rm i}^{0.22}\rho_{\rm t}^{-1}v^{1.3}g^{-0.66}\mathrm{sin}^{1.3}\theta m^{0.78},
\label{Vxcav}
\end{eqnarray}
where $\rho_{\rm i}$ and $\rho_{\rm t}$ are the densities of the impactor and the crust, and $v$, $\theta$, $L$, $m$, and $g$ are the impact speed and angle, impactor's diameter and mass, and gravity, respectively. Eq. (\ref{Vxcav}) is estimated from impact experiments \citep[e.g.,][]{sch87} and is consistent with recent computer simulations. This volume includes the crust not excavated but displaced, and the depth of the transient crater would be deeper than the thickness of Hadean continental crust. So, the excavated crust volume would be overestimated in this work.

According to \citet{abr12}, the diameter of the transient crater $D_{\rm t}$ is
\begin{eqnarray}
D_{\rm t}\!\!\!\!&=&\!\!\!\!1.44\left(\frac{gL}{v^2}\right)^{-0.22}\biggl(\frac{\rho_{\rm t}}{m}\biggr)^{-1/3} \nonumber \\
&=&\!\!\!\!1.37\rho_{\rm i}^{0.22/3}\rho_{\rm t}^{-1/3}v^{0.44}g^{-0.22}m^{0.26}.
\label{Dtm}
\end{eqnarray}
So the direct excavation (circular) area on the Earth's surface (i.e. horizontal cross section of the direct excavation region) is
\begin{eqnarray}
S_{\rm xcav}\!\!\!\!&=&\!\!\!\!1.63\left(\frac{gL}{v^2}\right)^{-0.44}\biggl(\frac{\rho_{\rm t}}{m}\biggr)^{-2/3} \nonumber \\
&=&\!\!\!\!1.48\rho_{\rm i}^{0.44/3}\rho_{\rm t}^{-2/3}v^{0.88}g^{-0.44}m^{0.52}.
\label{Sxcav}
\end{eqnarray}

The direct melting volume is
\begin{equation}
V_{\rm melt}=0.42\left(\frac{v^{2}}{\epsilon_{\rm m}}\right)^{0.84}\left(\frac{m}{\rho_{\rm t}}\right)\mathrm{sin}^{1.3}\theta,
\label{Vmelt}
\end{equation}
where $\epsilon_{m}$ is the specific internal energy of the target \citep{abr12}. The direct melting volume is proportional to the impactors' mass. When the depth of the melting region is deeper than the thickness of the crust, $h$, we use the following equation in place of Eq. (\ref{Vmelt}).
\begin{equation}
V_{\rm melt,h}=\pi\left\{\left(\dfrac{3V_{\rm melt}}{2\pi}\right)^{2/3}h-\dfrac{h^{3}}{3}\right\}.
\label{Vmelth}
\end{equation}
The melting region's shape is considered to be a hemisphere. From Eq. (\ref{Vmelt}), the direct melting area on the Earth's surface (i.e. horizontal cross section of the direct melting region) is
\begin{equation}
S_{\rm melt}=\pi\left(\dfrac{3V_{\rm melt}}{2\pi}\right)^{2/3}=1.08\left( \frac{v^{2}}{\epsilon_{\rm m}}\right)^{0.56}\left(\frac{m}{\rho_{\rm t}}\right)^{2/3}\mathrm{sin}^{2.6/3}\theta.
\label{Smelt}
\end{equation}
Equations (\ref{Vmelt}), (\ref{Vmelth}) and (\ref{Smelt}) implicitly assume that the target crusts have no geothermal gradient and a homogeneous initial temperature of 0$^\circ$C.

Then, we consider the subsequent effects. As craters collapse from transient craters to final craters due to gravity, their diameters become larger. There is a relationship between these diameters \citep{abr12},
\begin{equation}
D_{\rm t}=(D_{\rm c}^{0.15}D_{\rm f}^{0.85}) /1.2,
\label{Dt}
\end{equation}
where $D_{\rm t}$, $D_{\rm f}$, and $D_{\rm c}$ are the transient crater's diameter, final crater's diameter, and critical diameter between simple and complex craters, respectively. The final crater's diameter and the excavation area including gravitational collapse are
\begin{equation}
D_{\rm f}=1.79\rho_{\rm i}^{0.22/2.55}\rho_{\rm t}^{-0.1/2.55}v^{0.44/0.85}g^{-0.22/0.85}D_{\rm c}^{-0.15/0.85}m^{0.26/0.85},
\label{Dfm}
\end{equation}
\begin{equation}
S_{\rm xcav,f}=2.52\rho_{\rm i}^{0.44/2.55}\rho_{\rm t}^{-4/2.55}v^{0.88/0.85}g^{-0.44/0.85}D_{\rm c }^{-0.3/0.85}m^{0.52/0.85}.
\label{Sxcavf}
\end{equation}

Excavated or uplifting melts spread around the final crater \citep{abr13, osi11}. We assume the melting area including the area covered by melts, $S_{\rm melt,f}$, is equal to $S_{\rm xcav,f}$. When the impactor is larger than 100 km, the melting area including the area covered by melts can be expressed as the following equation,
\begin{eqnarray}
S_{\rm melt,f}\!\!\!\!&=&\!\!\!\!\frac{\pi}{4}f^{2}L^{2} \nonumber \\
&=&\!\!\!\!1.21f^{2}\left(\frac{m}{\rho_{\rm i}}\right)^{2/3},
\label{Smeltf}
\end{eqnarray}
where $f$ is the proportion of the diameter of the melt region to that of the impactor. The effects of decompression and adiabatic melting are included, and $f$ reaches 20--30 \citep{mar14}.

\subsection{Integrating the effects of a single impact} \label{Integrate}

The definition of $N_{\rm sfd}(>m_{\rm max})$ is
\begin{eqnarray}
N_{\rm sfd}(>m_{\rm max})\!\!\!\!&=&\!\!\!\!\frac{A}{\alpha-1}m_{\rm max}^{1-\alpha} \nonumber \\
&\!\!\!\!\approx&1,
\label{N(mmax)}
\end{eqnarray}
then,
\begin{equation}
A\approx (\alpha-1)m_{\rm max}^{\alpha-1},
\label{Ammax}
\end{equation}
\begin{equation}
m_{\rm max}\approx \left(\frac{A}{\alpha-1}\right)^{1/(\alpha-1)},
\label{mmaxA}
\end{equation}
where $m_{\rm max}$ is the maximum mass of the impactors \citep{zah97}. This $m_{\rm max}$ just represents the maximum mass value with the highest possibility, so it would have larger or smaller values actually.

The ratio of the collision probability with Earth and the moon is 23:1 \citep{zah97, ito06}. The proportional constants $A$ have the following relationship,
\begin{equation}
A_{\rm e}=23A_{\rm m},
\label{AeAm}
\end{equation}
where $A_{\rm e}$ and $A_{\rm m}$ are the $A$ values for Earth and the moon, respectively.

The effects of all LHB impacts are obtained by calculating the integral from $m_{\rm min}$ to $m_{\rm max}$ of the effects of a single impact. We describe all of the effects of a single impact (Eqs. (\ref{Vxcav}),  (\ref{Sxcav}), (\ref{Vmelt}), (\ref{Smelt}), (\ref{Sxcavf}), and (\ref{Smeltf})) as one general equation form, $I_{\rm vs}$, in order to easily understand their characteristics,
\begin{equation}
I_{\rm vs}=B_{j}m^{b_{j}},
\label{Ivs}
\end{equation} 
where $b_{j}$ is the power-law index of each $m$, and $B_{j}$ is a constant fixed by impact velocity and density (see Table \ref{tab1}). The general form of the effects of all LHB impacts is
\begin{eqnarray}
I_{\rm vs,T}\!\!\!\!&=&\!\!\!\!\int_{m_{\rm min}}^{m_{\rm max}}\frac{\mathrm{d} N_{\rm sfd}}{\mathrm{d}m}I_{\rm vs}dm \nonumber \\
&=&\!\!\!\!B_{j}\frac{A_{\rm e}}{1+b_{j}-\alpha}(m_{\rm max}^{1+b_{j}-\alpha}-m_{\rm min}^{1+b_{j}-\alpha}).
\label{IvsT}
\end{eqnarray}
When $m_{\rm min}$ is small enough,
\begin{equation}
I_{\rm vs,T}\approx\begin{cases}
    B_{j}\dfrac{A_{\rm e}}{1+b_{j}-\alpha}m_{\rm max}^{1+b_{j}-\alpha} & (0<\alpha<1+b_{j}) \\
    B_{j}\dfrac{A_{\rm e}}{\alpha-b_{j}-1}m_{\rm min}^{1+b_{j}-\alpha} & (1+b_{j}<\alpha). \\
\end{cases}
\label{IvsTapp}
\end{equation}
The total effects of the impacts are dependent on the impactor's maximum mass $m_{\rm max}$ when $0<\alpha<1+b_{j}$, and on the impactor's minimum mass $m_{\rm min}$ when $1+b_{j}<\alpha$. We only use Eq. (\ref{IvsTapp}) to understand the dependence of the effects on the impactor's mass, and always use Eq. (\ref{IvsT}) to estimate the scales of the effects.

The total melting volume is dependent on the impactor's maximum mass, $m_{\rm max}$ (when $\alpha<2$) and is proportional to the total mass, $M_{\rm T}$ (see Eq. (\ref{MT1})). The total direct melting area is proportional to the total cross section of the impactors, and the total cross section is proportional to $m^{5/3-\alpha}$. It has often been stressed that the effects of large impacts are greater than those of small impacts \citep[e.g.,][]{abr13}. However, we found that large impacts and their effects are important only when $\alpha$ is small, and small impacts are more important than large impacts when $1.52<\alpha$ and $5/3<\alpha$ because the number of such small impacts is very large. In this case, thin melt sheets formed by small impacts would have covered the entire surface of the Earth, and it can explain the absence of Hadean rocks (see Section \ref{absence}).

Especially when $\alpha$ is large, the total melting volume and area are strongly dependent on the minimum mass, $m_{\rm min}$. In this work, $m_{\rm min}$ is determined from the minimum mass $\mu_{\rm e}$ which can survive a fall through the Earth's atmosphere (see Section \ref{validity}). Assuming the current atmosphere, $\mu_{\rm e}=10^{11.5}$ g \citep{bla03}.
\begin{table}[htbp]
\caption{Coefficients}
\begin{center}
\begin{tabular}{lcc} \hline
Effects & $1+b_{j} $ & $B_{j} $ \\ \hline \hline
$V_{\rm xcav}$ & 1.78 & $0.127\rho_{\rm i}^{0.22}\rho_{\rm t}^{-1}v^{1.3}g^{-0.66}\mathrm{sin}^{1.3}\theta$ \\
$V_{\rm melt}$ & 2 & $0.42\rho_{\rm t}^{-1}\left(\dfrac{v^{2}}{\epsilon_{m}}\right)^{0.84}\mathrm{sin}^{1.3}\theta$ \\
$S_{\rm xcav}$ & 1.52 & $1.48\rho_{\rm i}^{0.44/3}\rho_{\rm t}^{2/3}v^{0.88}g^{-0.44}$ \\
$S_{\rm melt}$ & 5/3 & $1.08\rho_{\rm t}^{-2/3}\left(\dfrac{v^{2}}{\epsilon_{m}}\right)^{0.56}\mathrm{sin}^{2.6/3}\theta$ \\
$S_{\rm xcav,f}$, $S_{\rm melt,f}$ & 1.61 & $2.52\rho_{\rm i}^{0.17}\rho_{\rm t}^{-0.78}v^{1.04}g^{-0.52}D_{\rm c}^{-0.35}$ \\ 
$S_{\rm melt,f}$ ($L>100$ km) & 5/3 & $1.21f^{2}\rho_{\rm i}^{-2/3}$ \\ \hline
\end{tabular}
\end{center}
\label{tab1}
\end{table} 

These estimates do not include overlapping of craters. A better estimate including overlappings for the total excavation/melting area is expressed in the following equation \citep{fre80},  
\begin{equation}
S_{\rm r}=1-\mathrm{exp}\left(-\frac{S_{\rm T}}{S_{\rm e}}\right),
\label{Sr}
\end{equation}
where $S_{\rm T}$ is $S_{\rm xcav,T}$, $S_{\rm melt,T}$, $S_{\rm xcav,f,T}$ or $S_{\rm melt,f,T}$, and $S_{\rm e}$ is the total surface area of Earth. We use this correction in this paper. On the other hand, estimating the excavation/melting volumes including overlapping is difficult, so we do not use any corrections for them and they are probably overestimated.

\section{Effects of LHB impacts} \label{effects}

\subsection{Effects of LHB impacts with fixed total impactors' mass} \label{fixtotalmass}

We estimate the effects of LHB and their dependence on $\alpha$, where $M_{\rm T}$, the total mass of impactors, is fixed. We assume the $\alpha$ to be $1<\alpha<2$. From Eq. (\ref{dN(m)}), $M_{\rm T}$ is 
\begin{eqnarray}
M_{\rm T}\!\!\!\!&=&\!\!\!\!\int_{m_{\rm min}}^{m_{\rm max}} \frac{\mathrm{d} N_{\rm sfd}}{\mathrm{d}m}mdm \nonumber \\
&=&\!\!\!\!\frac{A}{2-\alpha}(m_{\rm max}^{2-\alpha}-m_{\rm min}^{2-\alpha}).
\label{MT1}
\end{eqnarray}
Then,
\begin{equation}
M_{\rm T}\approx\frac{Am_{\rm max}^{2-\alpha}}{2-\alpha},
\label{MT}
\end{equation}
where $m_{\rm min}$ is  small enough and $\alpha<2$ \citep{doh69}. This equation shows that the total mass depends on the maximum mass. The following equations derived from Eqs. (\ref{Ammax}) and (\ref{mmaxA}),
\begin{equation}
m_{\rm max} \approx \frac{2-\alpha}{\alpha-1}M_{\rm T},
\label{mmaxMT}
\end{equation}
\begin{equation}A=(2-\alpha)^{\alpha-1}(\alpha-1)^{2-\alpha}M_{\rm T}^{\alpha-1}.
\label{AeMT}
\end{equation}
The dependence of the effects on $M_{\rm T}$ is derived from Eqs. (\ref{IvsT}), (\ref{mmaxMT}) and (\ref{AeMT}),
\begin{equation}
I_{\rm vs,T}=B_{j}\frac{(2-\alpha)^{\alpha-1}(\alpha-1)^{2-\alpha}}{1+b_{j}-\alpha} \left\{\left(\dfrac{2-\alpha}{\alpha-1}M_{\rm T}\right)^{1+b_{j}-\alpha}-\mu_{\rm e}^{1+b_{j}-\alpha}\right\} M_{\rm T}^{\alpha-1}.
\label{IvsMT}
\end{equation} 

The most typical value of $M_{\rm T}$ is about 1--5$\times10^{23}$ g \citep[e.g.,][]{gom05, jor09, lev01}. \citet{abr13} also used the value $M_{\rm T}=2\times10^{23}$ g. In this paper, we use the value $M_{\rm T}=$ 1, 2 and $5\times10^{23}$ g to calculate the effects of LHB.

\begin{figure}[tbsp]
\begin{center}
\includegraphics[width=11cm]{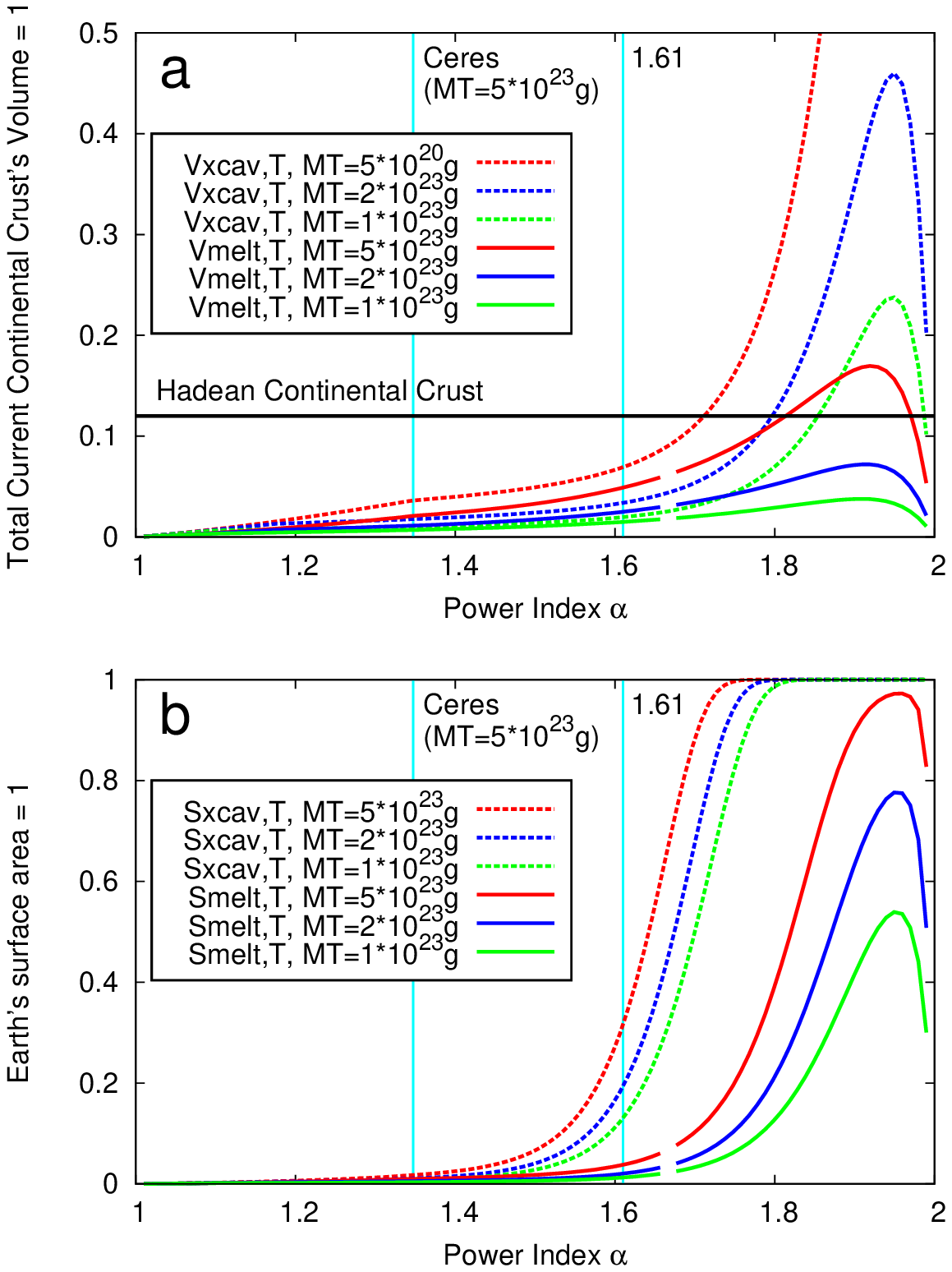}
\caption{Total direct effects of LHB with fixed total impactors' mass \newline Panel (a) and (b) show the total direct excavation/melting volumes and areas, respectively. Green, blue, and red curves show the effects when $M_{\rm T}=1$, 2, and $5\times10^{23}$ g, respectively. Solid curves show the total melting volumes and areas, dashed curves are those of excavation. The black line in panel (a) shows the total Hadean continental crust's volume. We cut off the SFD larger than the size of Ceres when $\alpha<1.35$ ($M_{\rm T}=5\times10^{23}$ g), left side of the aqua vertical line. In panel (a), the solid curves (i.e., melting volumes) include the correction of the thickness of the crust, $h=30$ km, but the dashed curves  (i.e., excavation volumes) do not.} \label{fig3}
\end{center}
\end{figure}

Figure \ref{fig3} (a) shows the estimated total excavation and melting volumes relative to the total Hadean continental crust's volume. We considered the thickness of the crust (30 km) for the estimates of the melting volume (see Section \ref{singlei}). We also consider the cut-off of the heavier side of the SFD to avoid the mass becoming unrealistically large. We chose the cut-off value as $m_{\rm ceres}=9.4\times10^{23}$ g, the mass of Ceres. Ceres is the largest object among the main belt asteroids. For example, when $M_{\rm T}=5\times10^{23}$ g, we have to cut off the SFD where $\alpha<1.35$. In this $\alpha$ region, $m_{\rm e,max}= m_{\rm ceres}$. The estimated total excavation/melting volumes are normalized with the total current continental crust's volume, $7.18\times10^{9}$ km$^{3}$ \citep{cog84}. According to geochemical constrains of the mantle, we assumed that the continental crust was formed at 4.5 Ga for the first time and there was about 12\% of the total current continental crust's volume in the last phase of the Hadean \citep{mcc94}. We assumed $\rho_{\rm i}=2.6$ g/cm$^{3}$, $\rho_{\rm t}=2.7$ g/cm$^{3}$, $v=21$ km/s and $\epsilon_{\rm m}=5.2$ MJ/kg. In the case of $M_{\rm T}=2\times10^{23}$ g in Fig. \ref{fig3} (a), the total melting volume is smaller than the total volume of Hadean continental crust in all $\alpha$ ranges. LHB impacts do not melt all of Hadean continental crust in this case. On the other hand, the total excavation volume exceeds that of the continental crust when $\alpha$ is larger than about 1.8, though we note that excavation volume of the continental crust would be overestimated (see Section \ref{singlei}). LHB can not excavate/melt all of the continental crust by direct effects when $\alpha$ =1.61.

Figure \ref{fig3} (b) shows the direct excavation/melting area using Table \ref{tab1}, and Eqs. (\ref{Sr}), (\ref{AeMT}), and (\ref{IvsMT}). This figure shows that LHB does not excavate/melt all of the Earth's surface area by direct effects when $\alpha$ =1.61. If there were many small impacts (i.e., when $\alpha$ was large), the excavation area would be able to expand the whole surface of the Earth.

\begin{figure}[tbsp]
\begin{center} 
\includegraphics[width=11cm]{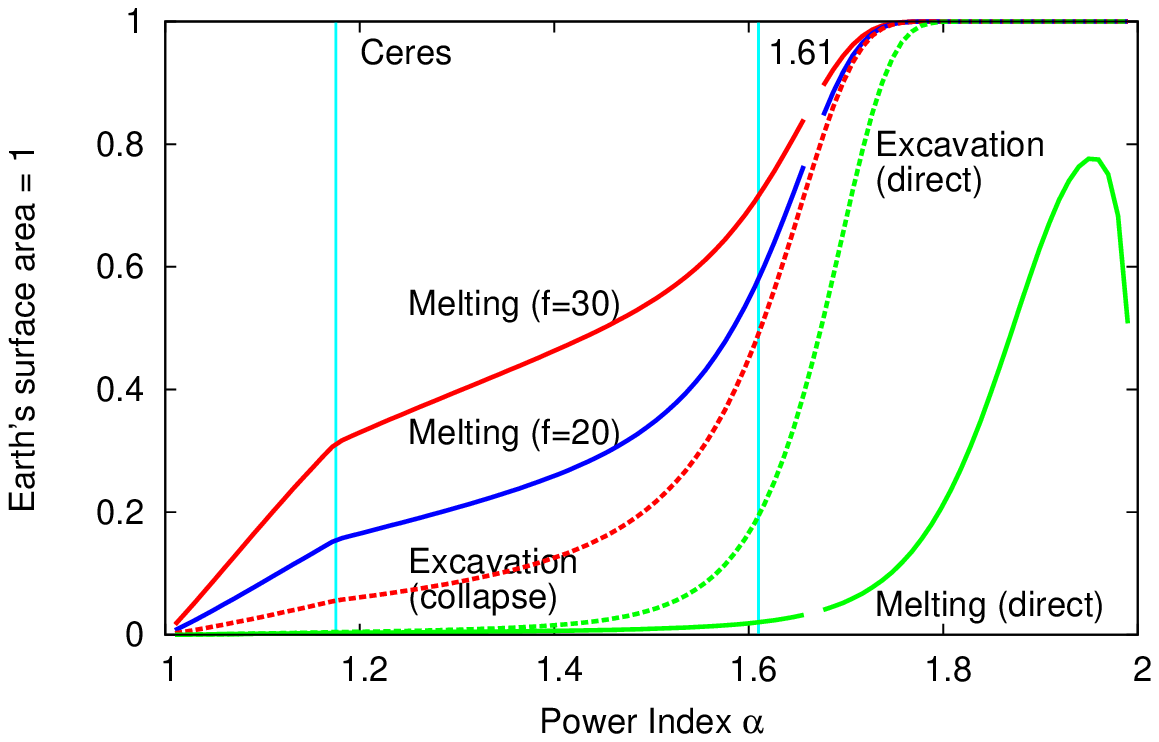}
\caption{Total excavation/melting areas with fixed total impactors' mass \newline Dashed and solid green curve show the direct excavation and melting areas, respectively. Red and blue curves show the total melting area including the area covered by melts when $f=30$ and 20, respectively. Dashed red curve shows the total excavation area including gravitational collapse. The dashed and solid green curves are consistent with dashed and solid blue curves in Fig. \ref{fig3} (b), respectively. Including the area covered by melts, the total melting area increases dramatically. The total mass that hit Earth, $M_{\rm T}$, is fixed at $2\times10^{23}$ g. We cut off the SFD larger than the size of Ceres when $\alpha<1.18$, left side of the aqua vertical line.} \label{fig4}
\end{center}
\end{figure}

Figure \ref{fig4} shows the total excavation/melting area including gravitational collapse and the area covered by melts. In particular, the total melting area including the area covered by melts (red and blue solid curves) is dramatically increased compared to the total direct melting area and approaches 60--70\% of the Earth's surface area where $\alpha=1.61$ and $f=$20--30 \citep{mar14}. As $\alpha$ becomes larger, the number of small impactors dramatically increases. For small impactors, the area covered by melts is comparable to the excavated area on the Earth's surface including gravitational collapse (Table \ref{tab1}). Therefore, the total melting area including the area covered by melts (the red and blue solid curves) approaches the total excavation area including gravitational collapse (the red dashed curve) when $\alpha$ is large enough.

\begin{figure}[tbsp]
\begin{center}
\includegraphics[width=11cm]{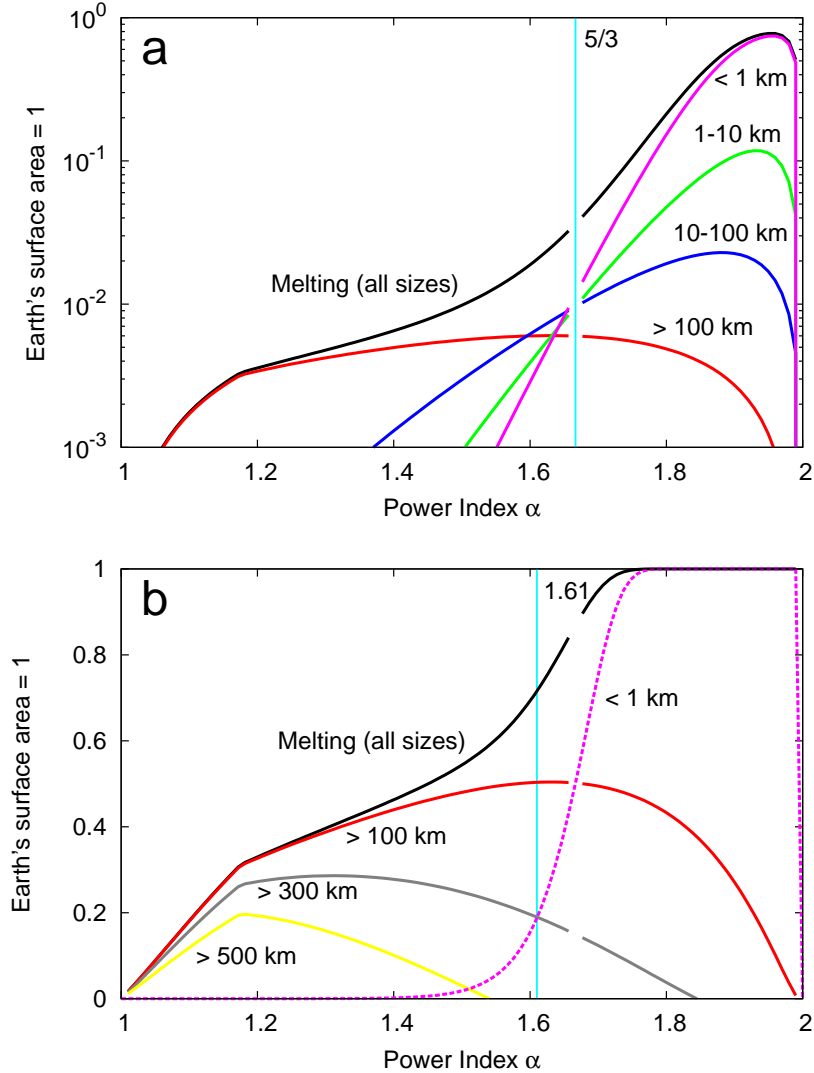}
\caption{Total melting areas for each impactor's size with fixed total impactor's mass \newline Panel (a) shows the total direct melting area; panel (b) shows the total melting area where $f=30$. Black curves in panels (a) and (b) show the total melting areas of all size impactors, consistent with the green and red solid curves in Fig. \ref{fig4}, respectively. In panel (b), red, gray and yellow curves show the total melting areas by impactors larger than 100, 300 and 500 km, respectively. Dashed pink curve shows the total direct melting area by impactors smaller than 1 km. $M_{\rm T}$ is fixed at  $2\times10^{23}$ g.} \label{fig5}
\end{center}
\end{figure}

Figure \ref{fig5} shows the total melting area for each impactor's size. Figure \ref{fig5} (a) shows the total direct melting area. It is mainly dependent on the maximum mass when $\alpha>5/3$ and on the minimum mass when $\alpha<5/3$ (also see $S_{\rm melt}$ in Table \ref{tab1}). In other words, as $\alpha$ becomes larger, the number of small impactors dramatically increases, and as $\alpha$ becomes smaller, the maximum mass to hit Earth increases. This trend is the same as that of the total impactors' cross sections. Also, even if we estimate the total direct melting area from the size of the maximum LHB basin on the moon or the lunar crater density, their dependence on $m$ is not changed, so that this trend of the total melting area is not changed (see Section \ref{lunarevidences}).

When the area covered by melts is included, this trend is changed. Figure \ref{fig5} (b) suggests that the trend of the contributions can be divided into three ranges. Small impactors mainly melt the Earth's surface when $\alpha$ is larger than about 1.7. Impactors of about 100 km in size mainly melt the surface when $\alpha$ is about 1.5--1.7. Impactors of over 500 km in size mainly melt the surface when $\alpha$ is less than about 1.3. Considering the ``stochastic'' view, an impactor of such size is the largest one in most cases. When $\alpha=1.61$, the effects of impactors of about 100 km in size are dominant.

\subsection{Estimate from lunar constraints} \label{lunarevidences}

While constraints of LHB on Earth have been erased, they remain on the moon. We estimate the value of $A_{\rm m}$ from lunar constraints when $1<\alpha<2$. First, we estimate it based on the largest basin formed by LHB impacts. The following equation is derived from Eqs. (\ref{Ammax}), where $m_{\rm m, max}$ is the maximum mass to hit the moon,
\begin{equation}
A_{\rm m}=(\alpha-1)m_{\rm m, max}^{\alpha-1} \label{aemmmax}.
\end{equation}
Because the impactor formed the Imbrium basin is at the edge of the SFD where the statistical fluctuation is the largest, this estimate has uncertainty. 

We assumed the largest impactor ($m_{\rm m, max}$) formed the Imbrium basin at 3.85 Ga. Although the South Pole-Aitken is the largest lunar basin, it was considered to be formed before the onset of LHB. The impactor's mass $m$ is estimated from $D_{\rm t}$ (Eq. (\ref{Dfm})),
\begin{equation}
m=0.149\rho_{\rm i}^{-0.11/0.35}\rho_{\rm t}^{1/0.78}v^{-0.22/0.13}g^{0.11/0.13}D_{\rm c}^{0.15/0.26}D_{\rm f}^{0.85/0.26}.
\label{m}
\end{equation}
Using this equation, the mass of the impactor formed the Imbrium is $m_{\rm m, max}=8.02\times10^{20}$ g, where $\rho_{\rm i}=2.6$ g/cm$^{3}$, $\rho_{\rm t}=$2.9 g/cm$^{3}$, $v=$18 km/s, $D_{\rm c}=$18 km on the moon, and the Imbrium basin's diameter, $D_{\rm f}$, is 1160 km \citep{spu93}.

Then, we estimate the value of $A_{\rm m}$ from the crater density on the moon. According to \citet{mor12}, the following differential equation represents the number of impacts to hit the moon:
\begin{equation}
\frac{\mathrm{d} N_{20}}{\mathrm{d}t}=2.7\times10^{-16}\mathrm{exp}(6.93t)+5.9\times10^{-7},
\label{NI94Sawtooth}
\end{equation}
where $N_{20}$ is the lunar crater density whose diameters are larger than 20 km, and $t$ is the age. The units are [$/\rm km^{2}$] and [Ga (Gyr ago)]. In this expression, LHB started at 4.1 Ga (a sawtooth-like timeline).

The number of impacts to hit area $S$ between the age $t$ and $t+\mathrm{d}t$ forming craters larger than $D$ is
\begin{equation}
n_{t,t+\mathrm{d}t}(>D)=\frac{\mathrm{d} N_{D}}{\mathrm{d}t}S.
\label{ntD}
\end{equation}
Then, the proportional constant $A$ is derived from Eq. (\ref{N(m)}),
\begin{equation}
A_{t,t+\mathrm{d}t}=(\alpha-1)Sm_{D}^{\alpha-1}\frac{\mathrm{d}N_{D}}{\mathrm{d}t},
\label{At}
\end{equation}
where $m_{D}$ is the impactor's mass that forms a crater whose diameter is $D$.  The total number of impacts which hit the moon to form a crater larger than 20 km during LHB is
\begin{eqnarray}
N_{\rm m,20}\!\!\!\!&=&\!\!\!\!S_{\rm m}N_{20} \nonumber \\
&=&\!\!\!\!S_{\rm m}\int_{t_{\rm f}}^{t_{0}}\dfrac{\mathrm{d} N_{20}}{\mathrm{d}t}dt \nonumber \\
&=&\!\!\!\!3.79\times10^{7}\times8.76\times10^{-5} \nonumber \\
&=&\!\!\!\!3.32\times10^{3},
\label{Nm}
\end{eqnarray}
where $S_{\rm m}$ is the moon's surface area, $t_{0}=4.5$ and $t_{\rm f}=0$. This estimate of $N_{20}=8.76\times10^{-5}$ km$^{-2}$ is exactly consistent with the real lunar crater density in the Nectaris basin \citep{mar12}. The key point is that we use the timeline only for calculating the lunar crater density $N_{20}$ and so this estimate does not depend on a specific LHB model. Therefore, $A_{\rm m}$ can be directly estimated only by observing the lunar crater density. Then,
\begin{equation}
A_{\rm m}=(\alpha-1)m_{\rm m,20}^{\alpha-1}N_{\rm m,20},
\label{AmDensity}
\end{equation}
where $m_{\rm m, 20}$ is the impactor's mass which can make a 20-km-diameter crater on the moon. Using Eq. (\ref{m}), $m_{\rm m,20}=1.38\times10^{15}$ g. We summarize how to estimate $A_{\rm e}$ and $m_{\rm e, max}$ from lunar constraints in Table \ref{tab:values}.
\begin{table}[htbp]
\caption{How to calculate each value}
\begin{center}
\begin{tabular}{lcc} \hline
Constraints & $A_{\rm e}$ & $m_{\rm e, max}$ \\ \hline \hline
Total mass & $(2-\alpha)^{\alpha-1}(\alpha-1)^{2-\alpha}M_{\rm T}^{\alpha-1}$ & $\dfrac{2-\alpha}{\alpha-1}M_{\rm T}$ \\
Imbrium size & $23(\alpha-1)m_{\rm m,max}^{\alpha-1}$ & $23^{1/(\alpha-1)}m_{\rm m,max}$ \\
Crater density & $23(\alpha-1)m_{\rm m,20}^{\alpha-1}N_{\rm m,20}$ & $\left(23m_{\rm m,20}^{\alpha-1}N_{\rm m,20}\right)^{1/(\alpha-1)}$ \\ \hline
\end{tabular}
\end{center}
\label{tab:values}
\end{table}

\begin{figure}[tbsp]
\begin{center}
\includegraphics[width=11cm]{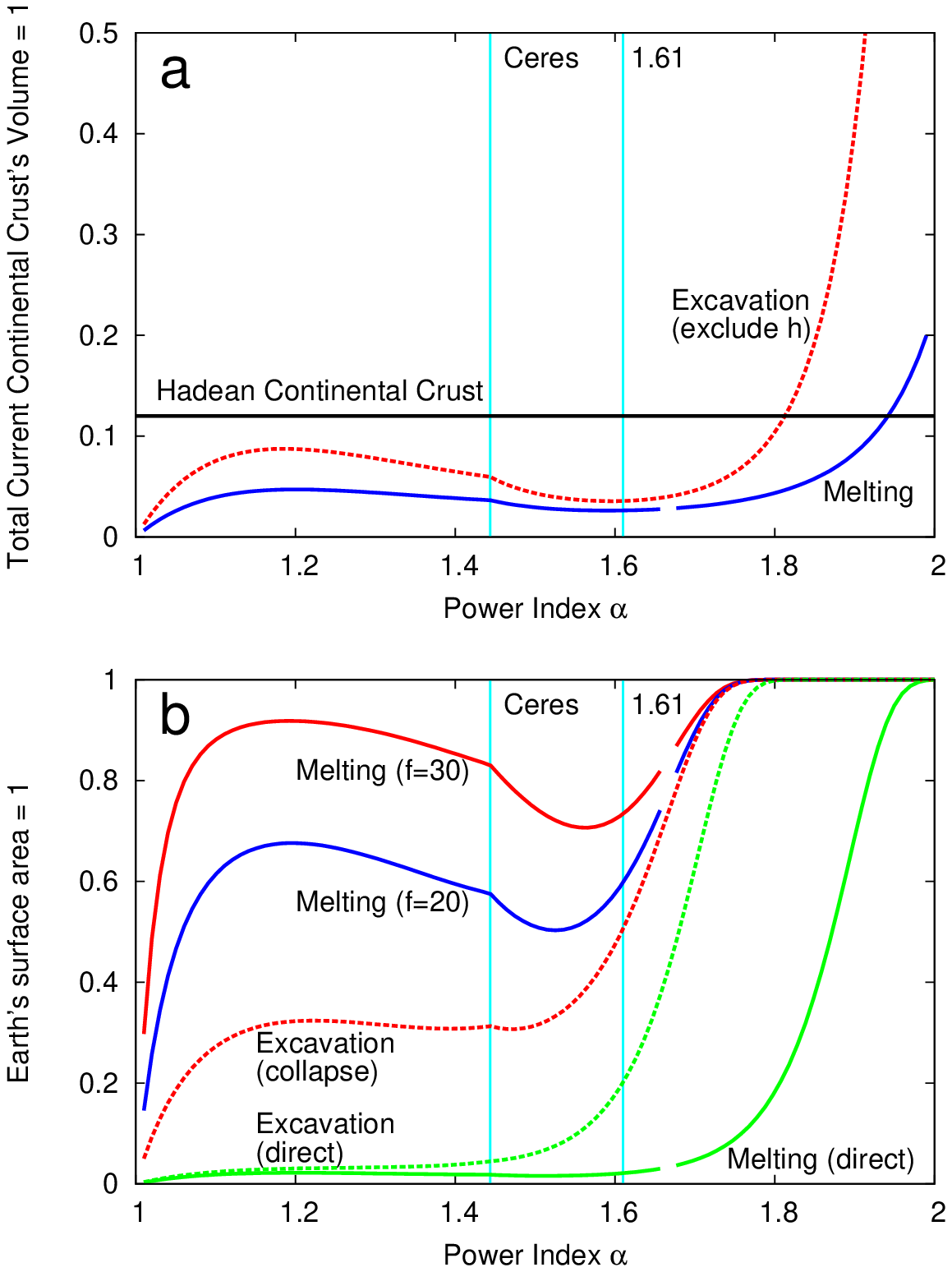}
\caption{Total effects of LHB estimated from the size of the Imbrium basin \newline Panel (a) and (b) show the total excavation/melting volumes and areas, respectively. In panel (a), red and blue curves show the total excavation and melting volumes, respectively. The black line shows the total Hadean continental crust's volume. The melting volume includes the correction of the thickness of the crust, $h=30$ km, but the excavation volume does not. In panel (b), dashed and solid green curves show the total direct excavation and melting areas, respectively. Dashed red curve shows the total excavation areas including gravitational collapse. Red and blue solid curves show the total melting areas including the area covered by melts where $f=30$ and 20, respectively. The Imbrium mass is $m_{\rm m, max}=8.02\times10^{20}$ g. We cut off the SFD larger than the size of Ceres when $\alpha<1.44$.} \label{fig6}
\end{center}
\end{figure}

\begin{figure}[tbsp]
\begin{center}
\includegraphics[width=11cm]{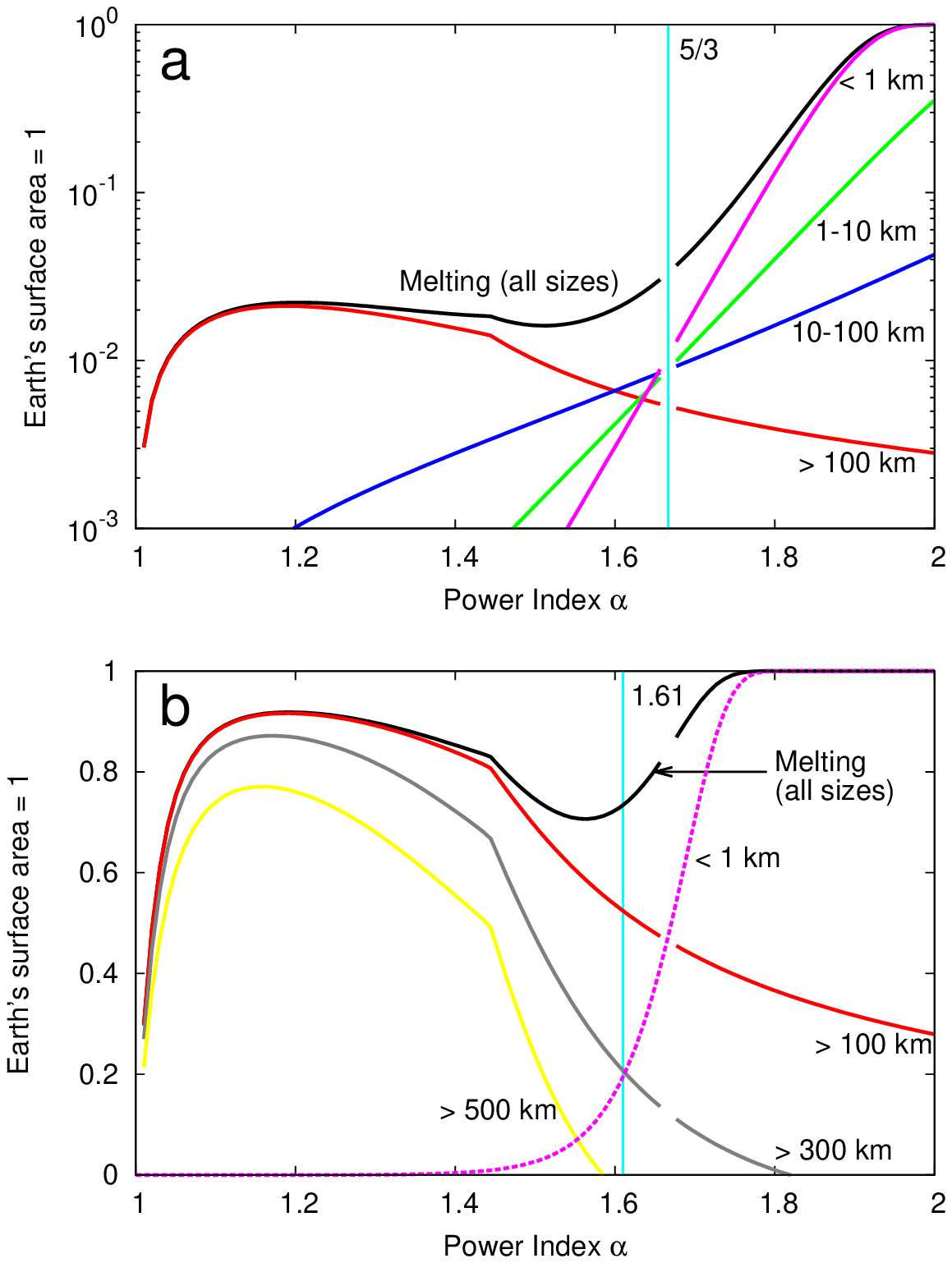}
\caption{Total melting areas for each impactor's size estimated from the size of the Imbrium basin \newline Panel (a) represents the direct total melting area; panel (b) shows the total melting area including the area covered by melts where $f=30$. Black curves in panels (a) and (b) show the total melting areas of all size impactors, consistent with the green and red solid curves in Fig. \ref{fig6} (b), respectively. In panel (b), red, gray, and yellow curves show the total melting areas by impactors larger than 100, 300, and 500 km, respectively. Dashed pink curve shows the total direct melting area by impactors smaller than 1 km. Imbrium mass is $m_{\rm m, max}=8.02\times10^{20}$ g.} \label{fig7}
\end{center}
\end{figure}

\begin{figure}[tbsp] \begin{center}
\includegraphics[width=11cm]{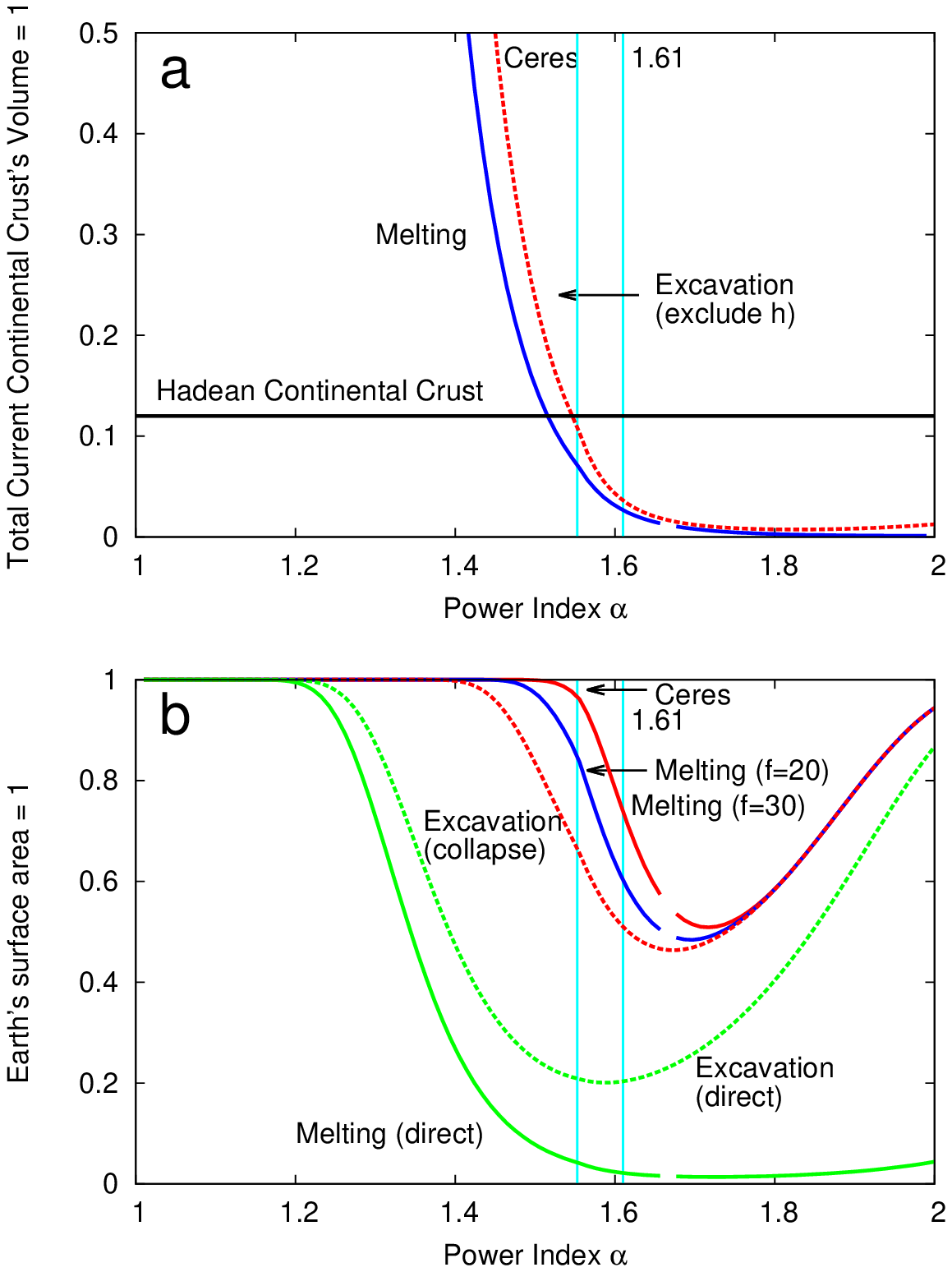}
\caption{Total effects of LHB estimated from the lunar crater density \newline Panel (a) and (b) show the total excavation/melting volumes and areas, respectively. In panel (a), red and blue curves show the total excavation and melting volumes, respectively. The black line shows the total Hadean continental crust's volume. The melting volume includes the correction of the thickness of the crust, $h=30$ km, but the excavation volume does not. In panel (b), dashed and solid green curves show the total direct excavation and melting areas, respectively. Dashed red curve shows the total excavation area including gravitational collapse. Red and blue solid curves show the total melting areas including the area covered by melts where $f=30$ and 20, respectively. The density of lunar craters larger than 20 km is $N_{20}=8.76\times10^{-5}$ km$^{-2}$. We cut off the SFD larger than the size of Ceres when $\alpha<1.55$.} \label{fig8}
\end{center}
\end{figure}

\begin{figure}[tbsp]
\begin{center}
\includegraphics[width=11cm]{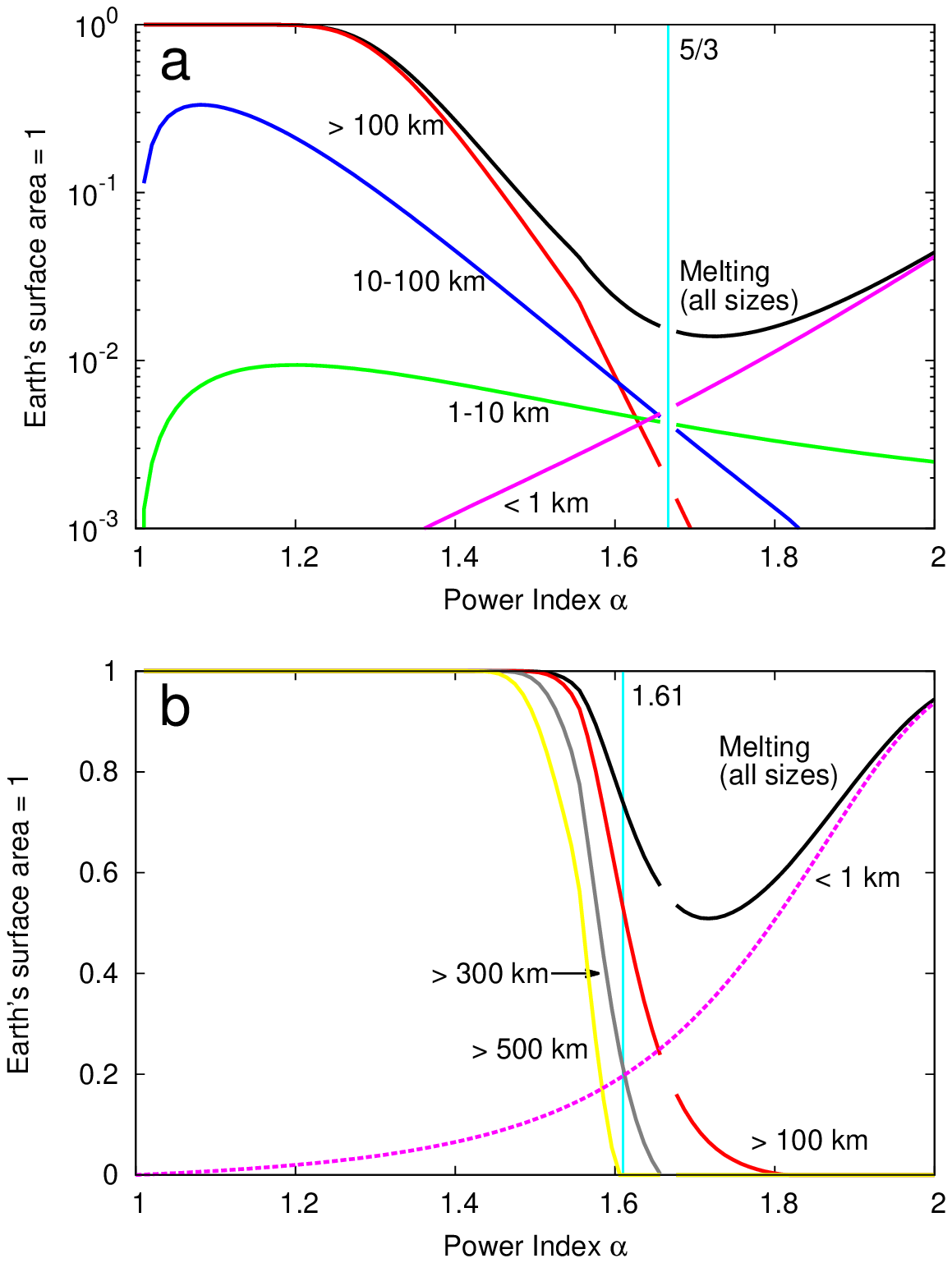}
\caption{Total melting ares for each impactor's size estimated from the lunar crater density \newline Panel (a) represents the total direct melting area; panel (b) shows the total melting area including the area covered by melts where $f=30$. Black curves in panels (a) and (b) show the total melting areas of all size impactors, consistent with the green and red solid curves in Fig. \ref{fig8} (b), respectively. In panel (b), red, gray, and yellow curves show the total melting areas by impactors larger than 100, 300, and 500 km, respectively. Dashed pink curve shows the direct melting rate by impactors smaller than 1 km. The density of lunar craters larger than 20 km is $N_{20}=8.76\times10^{-5}$ km$^{-2}$.} \label{fig9}
\end{center}
\end{figure}

Figures \ref{fig6} and \ref{fig7} show the total excavation/melting volume and area estimated from the size of the Imbrium. Fig. \ref{fig7} (b) includes the subsequent effects, while Fig. \ref{fig7} (a) does not. On the other hand, Figs. \ref{fig8} and \ref{fig9} show those estimated from the crater density on the moon. Fig. \ref{fig9} (b) includes the subsequent effects, while Fig. \ref{fig9} (a) does not.

Figures \ref{fig6} (a) and \ref{fig8} (a) show that LHB does not excavate/melt all of Hadean continental crust in almost all  $\alpha$ ranges. This is the same result as that of the excavation/melting volume estimated from the total mass to hit the Earth (Fig. \ref{fig3} (a)). Figures \ref{fig6} (b) and \ref{fig8} (b) show that, although LHB does not excavate/melt all of the Earth's surface directly, most of the surface is covered by the melts from the subsequent effect in almost all $\alpha$ ranges. Figures \ref{fig7} (a) and \ref{fig9} (a) show the total direct melting areas for each impactor's size. They show that the total melting areas are mainly dependent on the maximum mass when $\alpha>5/3$ and on the minimum mass when $\alpha<5/3$ (also see $S_{\rm melt}$ in Table \ref{tab1}). Figures \ref{fig7} (b) and \ref{fig9} (b) suggest that the trend in contributions to covering the surface with melts are divided into three $\alpha$ ranges like the contribution derived from the total mass (Fig. \ref{fig5} (b)). The three $\alpha$ ranges are about 1.0--1.5, 1.5--1.7, and 1.7--2.0 for both the estimates derived from the size of the Imbrium basin and the lunar crater density. When $\alpha=1.61$, the effects of impacts of about 100 km in size are dominant. 

\begin{figure}[tbsp]
\begin{center}
\includegraphics[width=11cm]{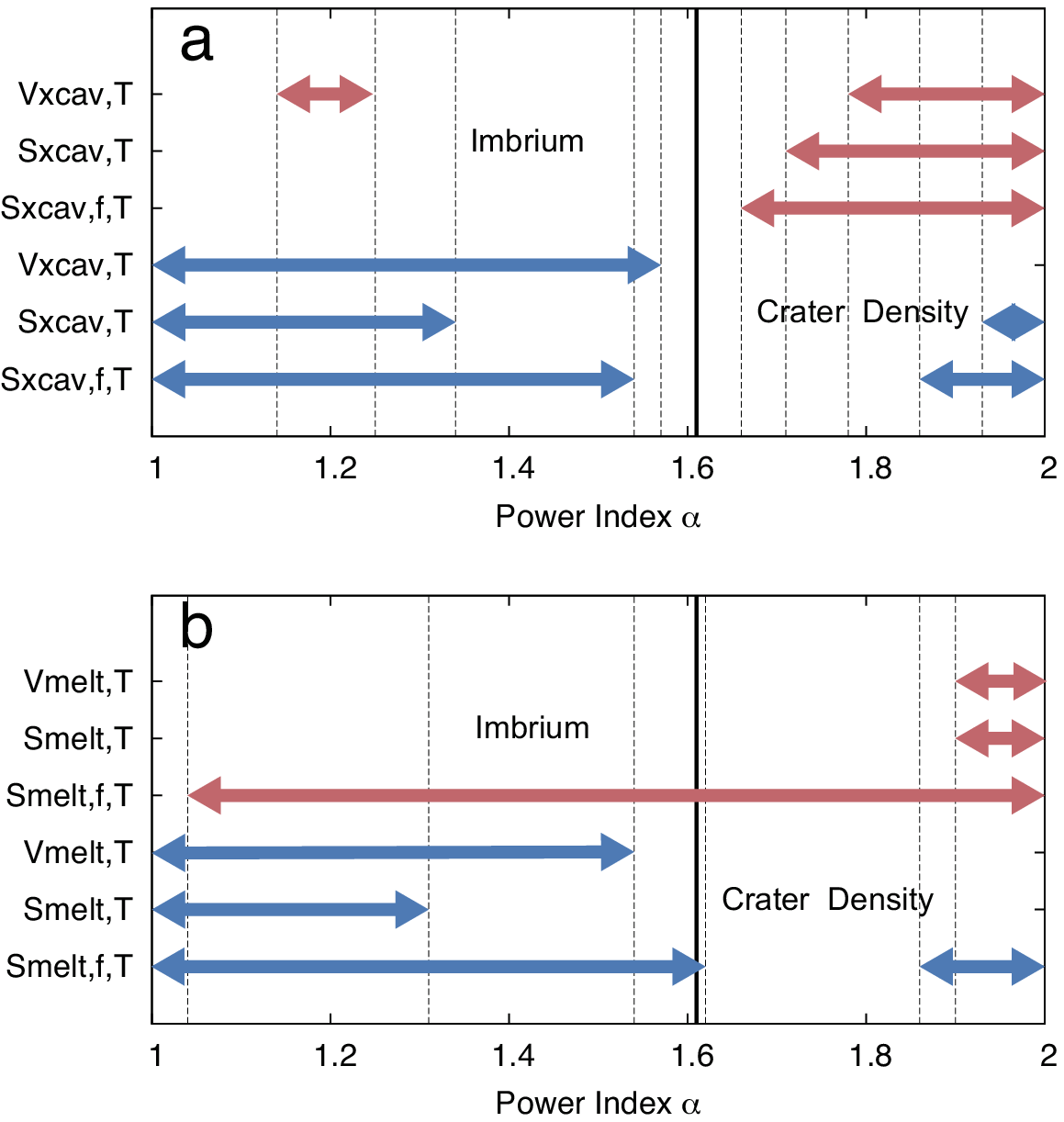}
\caption{The $\alpha$ ranges where LHB excavates (a) and melts (b) 70\% of total Hadean continental crust \newline Red arrows show the $\alpha$ ranges of LHB estimated from the size of Imbrium basin that can excavate/melt 70\% of the total Hadean continental crust's volume (i.e., 8.4\% of the total current continental crust's volume) or the Earth's surface area. Blue arrows show the $\alpha$ ranges estimated from the lunar crater density. Black solid lines show $\alpha=1.61$. In the estimate of $S_{\rm melt,f,T}$, we assumed $f=30$. When we consider the subsequent effects, the $\alpha$ ranges in which LHB can excavate/melt most of the continental crust expand.} \label{fig10}
\end{center}
\end{figure}

We summarize the above results in Fig. \ref{fig10}. The red and blue arrows represent the $\alpha$ ranges of LHB estimated from the size of the Imbrium basin and the lunar crater density that can excavate/melt 70\% of Hadean continental crust's volume or the Earth's surface area. Considering only the direct excavation/melting of the continental crust, it is difficult for LHB impacts for most $\alpha$ ranges to excavate/melt all of the continental crust. According to the estimate from the size of the Imbrium basin, there is enough excavation/melting only for limited conditions, when $\alpha>1.7$ (excavation) and $\alpha>1.9$ (melting), the $\alpha$ ranges where the total effects are enough for both the volume and area. According to the estimate from the lunar crater density, there is enough excavation and melting only for $\alpha<1.3$. When $\alpha=1.61$, all estimated direct effects of LHB impacts are not enough. On the other hand, when we consider the subsequent effects, the $\alpha$ ranges in which LHB can excavate/melt the continental crust expand. In particular, the subsequent melting can cover over 70\% of the Earth's surface in all $\alpha$ ranges (estimated from the size of the Imbrium) and when about $\alpha<1.6$ and $\alpha>1.9$ (estimated from the lunar crater density). When $\alpha=1.61$, the subsequent melting covers over 70\% of the surface in both the estimates. In conclusion, our results show that LHB would not excavate/melt all of the Hadean continental crust directly, but most of the Earth's surface area could be covered by melts.

\section{Discussion}

\subsection{LHB and the absence of Hadean rocks} \label{absence}

Can LHB explain the absence of Hadean rocks? When a crust is excavated by impacts, it is broken into small pieces and scattered. This helps Hadean continental crust to subduct with oceanic plates and disappear from the Earth's surface. When the crusts melt, their radiometric ages of zircons are reset \citep{abr13}. It erased the record of Hadean rocks. If the melts covered the entire Earth's surface area, and the stratigraphic succession has been preserved until today, the absence of Hadean rocks on the surface can be explained. Our result that LHB is not able to excavate/melt all of Hadean continental crust directly but can cover most of the Earth's surface with 3-km thick melts suggests that if there has not been a large-scale folding or overturn of the crust until today, LHB could explain why we have never found any Hadean rocks except the Jack Hills zircons because most of Hadean continental crust is not be exposed on the Earth's surface in this case.

\subsection{Comparison of results to previous works} \label{vsprevious}

We compare our results derived from the lunar constraints with previous works with $\alpha=1.61$. In Table \ref{tab3}, we summarized SFDs and the excavation/melting volumes and areas of our calculations, \citet{abr13} and \citet{mar14}. Without the subsequent effects, both our calculation and \citet{abr13} show that a few crusts are excavated/melted by LHB. Not considering crater overlapping, the total melting area including the area covered by melts, $S_{\rm melt,f, T}$, is over 90\%. This estimate is consistent with that of \citet{mar14}. However, if we take into account crater overlapping, the coverage is reduced to 60-74\%.
\begin{table}[htbp]
\caption{Comparison between our work and previous works}
\begin{center}
\begin{tabular}{lccc} \hline
 & Our work & \citet{abr13} & \citet{mar14} \\ \hline \hline
SFD & $\alpha$=1.61 & MBA\tablenotemark{a} & MBA\tablenotemark{a} \\
$v$ [km/s] & 21 & 20 & 25 \\
$\rho_{\rm i} \ \rm [g/cm^{3}$] & 2.6 & 2.7 & 3.314 \\
$\rho_{\rm t} \ \rm [g/cm^{3}$]  & 2.7 & 2.7 & "granite"  \\
$\mu_{\rm e}$ [g] & $10^{11.5}$ & $1.41\times10^{15}$\tablenotemark{b} & $5.86\times10^{18}$\tablenotemark{c}\\
$T_{\rm surf} \ \rm [C^{\circ}$] & 0 & 20 & 20 \\
d$T$/d$z$ [C$^{\circ}$\!\!/km] & 0 & 12, 70 & 11.25 \\
$h$ [km] & 30 & - & 30 \\ 
$M_{\rm T}$ [g] & $2.2\times10^{23}$ & $2\times10^{23}$ & - \\ \hline
$V_{\rm xcav,T}$ (\%)\tablenotemark{d} & 3.6 & $\sim$7\tablenotemark{e} & - \\
$V_{\rm melt,T}$ & 2.7 & $\sim$2.1-3.6\tablenotemark{f} & - \\
$S_{\rm xcav,T}$ (\%)\tablenotemark{g} & 20 & $\sim$25 & - \\
$S_{\rm melt,T}$ & 2.2 & $\sim$5-10 & - \\
$S_{\rm xcav,f,T}$ & 51 & - & - \\
$S_{\rm melt,f,T}$ & 60-74 & - & - \\
$S_{\rm melt,f,T}$\tablenotemark{h} & 92-130 & - & 70-100 \\ \hline
\end{tabular}
\tablenotetext{a}{Main belt asteroids' SFD}
\tablenotetext{b}{$L$=1km}
\tablenotetext{c}{$L$=15km}
\tablenotetext{d}{Percent of the total current continental crust's volume}
\tablenotetext{e}{``Impact ejecta covered the entire surface of the LHB-era Earth to a depth close to 1 km.''}
\tablenotetext{f}{``$\sim$1.5--2.5 vol.\% of the upper 20km of Earth's  crust was melted in the LHB.''}
\tablenotetext{g}{Percent of the Earth's surface area}
\tablenotetext{h}{Not considering crater ovarlapping}
\end{center}
\label{tab3}
\end{table} 

\begin{figure}[tbsp]
\begin{center} 
\includegraphics[width=11cm]{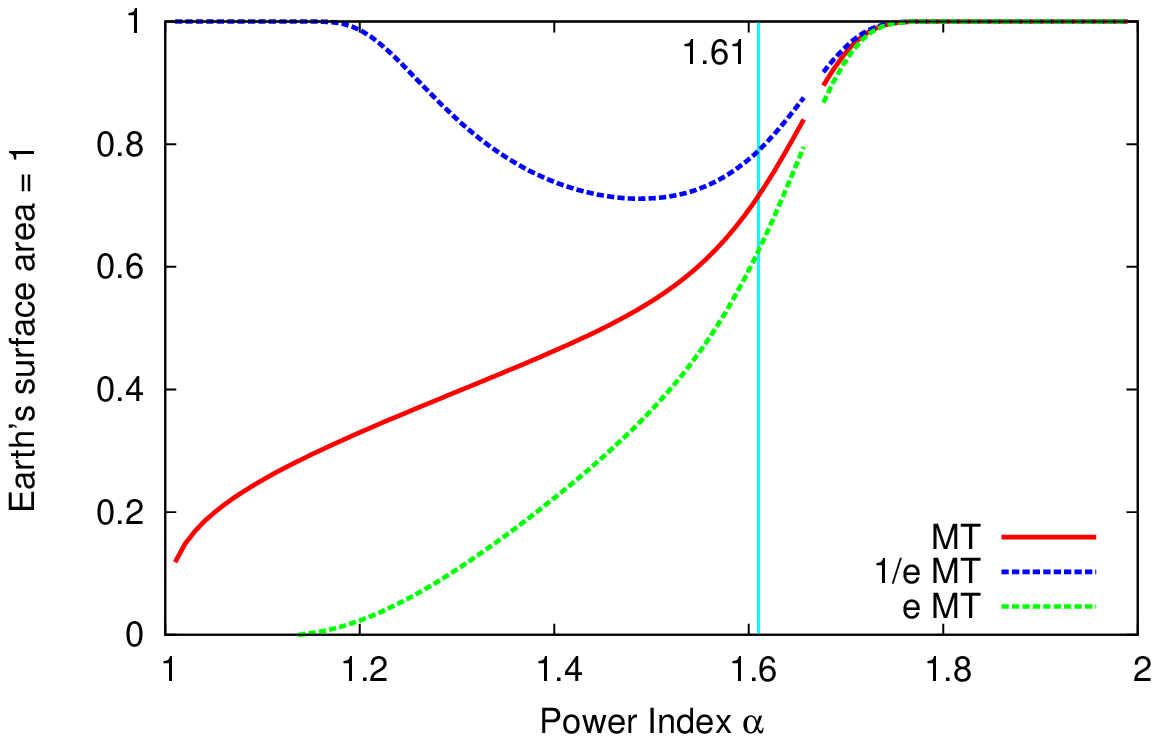}
\caption{Total melting areas in a stochastic scenario with fixed total impactors' mass \newline  Blue and green dashed curves show the total melting areas including the area covered by melts when the maximum masses are $1/e$ and $e$ times as heavy as the previous maximum mass, $m_{\rm e,max}$, respectively. Red solid curve shows our analytical estimate of the case with highest possibility. We did not consider the cut off of the SFD  by the mass of Ceres.} \label{fig11}
\end{center}
\end{figure}

Note that while we analytically estimated the values of LHB effects at the case of highest probability, \citet{mar14} adopted Monte Carlo simulations so that their results have a ``stochastic'' nature. To roughly take into account the stochastic nature in our calculations, we have the maximum mass fluctuate by a value from  $1/e$ to $e$ because the fluctuation of maximum mass impact has a large contribution to results. Figure \ref{fig11} shows the total melting area including the area covered by melts for $M_{\rm T}=2\times10^{23}$ g. When $\alpha$ is small, the fluctuation affects the melting rate significantly because the contribution from the maximum mass is larger for smaller $\alpha$.

\subsection{Validity of assumptions and models} \label{validity}

Here we discuss the validity of the assumptions we adopt. First, we did not consider the geothermal heat. We obtained $V_{\rm melt}$, that is consistent with the analytical shock-heating model used in \citet{abr13}, where the target has no geothermal gradient and the initial temperature is homogenized to be $0^{\circ}$C. Because geothermal heat increases the melting volume, $V_{\rm melt}$ that we have obtained could be an underestimate. If the Hadean Earth had a higher geothermal gradient than that of the present Earth, the melting volume could increase by a factor of two or three at most \citep{abr13}. On the other hand, the subsequent melting includes the gradient's effect. However, \citet{mar14} claimed that if the Hadean mantle potential temperature was higher or the lithosphere was thinner than today, the melting volume may increase by 50--75\%.

\begin{figure}[tbsp]
\begin{center} 
\includegraphics[width=11cm]{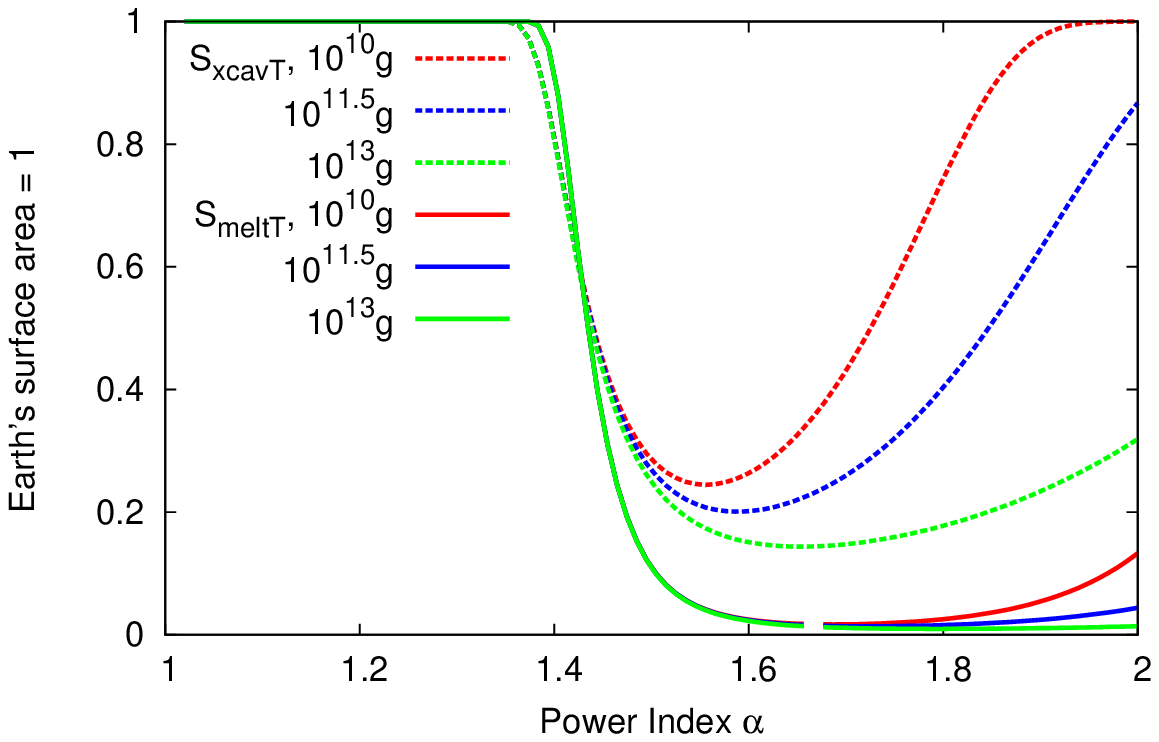}
\caption{Dependence of the total direct excavation/melting areas on $\mu_{\rm e}$ estimated from the lunar crater density \newline Red, blue and green curves show the total direct excavation/melting areas when $\mu_{\rm e}=10^{10}$, $10^{11.5}$ and $10^{13}$ g, respectively. Dashed and solid curves show the total direct excavation and melting areas, and blue ones are consistent with the dashed and solid green curves in Fig. \ref{fig8} (b), respectively. These estimates are very dependent on the minimum size of the impactor, especially when $\alpha$ is large.} \label{fig12}
\end{center}
\end{figure}

Second, our estimates of the minimum impactor mass may be overestimated because the Hadean atmosphere may have had higher pressure than the current atmosphere. The minimum diameter $L_{\rm min}$ can be shown analytically as
\begin{equation}
L_{\rm min}=0.15\frac{P_{\rm surf}}{\rho_{\rm i}g_{\rm surf}\rm sin\theta},
\label{Lmin}
\end{equation}
where $P_{\rm surf}$ and $g_{\rm surf}$ are, respectively, the atmospheric pressure at the surface and the acceleration of gravity \citep{mel89}. In this case, our estimates of the particular excavation/melting areas decrease when $\alpha$ is large (Fig. \ref{fig12}).

Third, our estimate assumes impactors' density, $\rho_{\rm i}$, of 2.6 g/cm$^{3}$ and velocity, $v$, of 21 km/s during LHB, and these values imply that the impactors were asteroids. However, some previous works claimed that the source of LHB impacts were comets or icy planetesimals \citep[e.g.,][]{lev01, jor09}. In these cases, the impactor's density should be changed to $\rho_{\rm i}\sim1$ g/cm$^{3}$ and the velocity to $v\sim30$ km/s. However, this change would not affect the results significantly. According to Table \ref{tab1}, $V_{\rm xcav}\propto \rho_{\rm i}^{0.22}v^{1.3}m^{0.78}$, $S_{\rm xcav}\propto \rho_{\rm i}^{0.44/3}v^{0.88}m^{0.52}$, $S_{\rm xcav,f}\propto \rho_{\rm i}^{0.17}v^{1.0}m^{0.61}$, $V_{\rm melt}\propto v^{1.68}m$ and $S_{\rm melt}\propto v^{1.12}m^{2/3}$, so these effects should increase by only $(1.0/2.6)^{0.22+0.78}\times (30/21)^{1.3}=0.61$, $(1.0/2.6)^{0.44/3+0.52}\times (30/21)^{0.88}=0.72$, $(1.0/2.6)^{0.17+0.61}\times (30/21)^{1.0}=0.68$, $(30/21)^{1.68}\times(1.0/2.6)=0.70$ and $(30/21)^{1.12}\times(1.0/2.6)^{2/3}=0.79$ times, respectively. $S_{\rm melt,f}$ does not depend on $\rho_{\rm i}$ and $v$. If the effect is regulated by the impactor's kinetic energy, $f$ should be multiplied $(1.0/2.6\times (30/21)^2)^{(1/3)}=0.92$ times, and the minimum size impactor that can form the flood melt should be multiplied $(1.0/2.6\times (30/21)^2)^{(-1/3)}=1.1$ times.

Finally, we discuss the effects of pre-LHB impacts. In the sections above, we only estimated the effects of LHB. However, several impacts occurred during the middle of the Hadean eon and much more impacts occurred at the beginning of it. In the first few hundred million years of the Hadean eon, impacts melted (or excavated) Hadean continental crust right after they formed, and the continental crust must not have had time to grow sufficiently.

\section{Summary}

We have investigated by analytical arguments the possibility for LHB impacts to excavate/melt Hadean continental crust. In order to reveal intrinsic physics, we adopt simple power-law impactors' SFD with various exponents $\alpha$, rather than a single detailed SFD. We divided the effects of impactors into two phases, and derived general formulas of excavation/melting volume and area as functions of $\alpha$ and the impactor's mass multiplied by a factor determined by impact velocity, planetary gravity, bulk density of impactors and the target planet. We estimated the total LHB effects from the total mass of LHB impacts and two types of constraints on the moon, the size of the largest basin during LHB and the small crater density.

With the fixed total LHB mass, the total direct melting area on the Earth's surface is generally regulated by small (large) impacts, for large (small) $\alpha$. The estimates from the lunar constraints suggest that LHB can excavate/melt almost all of Hadean continental crust in narrow $\alpha$ ranges. Estimating from the size of the Imbrium basin, LHB can remove the continental crust only at $\alpha > 1.7$ (excavation) and $\alpha > 1.9$ (melting), while estimating from the lunar crater density, only at $\alpha < 1.3$ (excavation and melt). In contrast, the subsequent melts which spread on and beyond the final craters can cover over 70\% of the Earth's surface in all $\alpha$ ranges (estimated from the size of the Imbrium basin) or when about $\alpha < 1.6$ (estimated from the lunar crater density). However, the most likely value of $\alpha$ is 1.6--1.7. We conclude that LHB impacts would not excavate/melt all of Hadean continental crust directly, but most of the Earth's surface could be covered by melts of subsequent impact effects. It suggests the absence of Hadean rocks could be explained by LHB if the stratigraphic succession has been preserved until today because most of Hadean continental crust is not be exposed on the Earth's surface in this case.

\acknowledgments

We thank S. Marchi and O. Abramov for very useful comments as reviewers. Their comments helped us improve the paper a lot. We thank S. Maruyama, K. Kurosawa, and H. Sawada for valuable discussions. T. S. was supported by a Grant-in-Aid for Young Scientists (B), JSPS KAKENHI Grant Number 24740120, and Grant-in-Aid for Scientific Research on Innovative Areas Number 2605,  MEXT. This research was supported by a grant for JSPS (23103005) Grant-in-aid for Scientific Research on Innovative Areas.

\appendix

\section{Value of $\alpha$ that fulfills both lunar constraints} \label{appendix}

We calculated the $\alpha$ value that fulfills both lunar constraints of the size of the Imbrium basin and the lunar crater density. From Table \ref{tab:values},
\begin{equation}
23(\alpha-1)m_{\rm m, max}^{\alpha-1} =23(\alpha-1)m_{\rm m,20}^{\alpha-1}N_{\rm m,20},
\label{AemmmaxDensity}
\end{equation}
thus
\begin{eqnarray}
\alpha\!\!\!\!&=&\!\!\!\!1+\dfrac{\mathrm{ln}N_{\rm m,20}} {\mathrm{ln}(m_{\rm m, max}/m_{\rm m,20})} \nonumber \\
&=&\!\!\!\!1.61,
\label{fulfillalpha}
\end{eqnarray}
where $N_{\rm m,20}=3.32\times10^{3}$, $m_{\rm m, max}=8.02\times10^{20}$ g, and $m_{\rm m,20}=1.38\times10^{15}$ g (see Section \ref{lunarevidences}).

\end{document}